# Coexisting 1T/2H polymorphs, reentrant resistivity behavior and charge distribution in MoS$_2$-hBN 2D/2D composite thin films

Swati Parmar,[1,2] Abhijit Biswas,[1] Sachin Kumar Singh,[1] Bishakha Ray,[3] Saurabh Parmar,[3] Suresh Gosavi,[4] Vasant Sathe,[5] Ram Janay Choudhary,[5] Suwarna Datar,[3]* and Satishchandra Ogale[1]*

[1]Department of Physics and Centre for Energy Science, Indian Institute of Science Education and Research (IISER) Pune, Maharashtra-411008, India
[2]Department of Technology, Savitribai Phule Pune University, Maharashtra-411007, India
[3]Defense Institute of Advanced Technology, Pune, Maharashtra-411025, India
[4]Department of Physics, Savitribai Phule Pune University, Maharashtra 411007, India
[5]UGC-DAE Consortium for Scientific Research, Indore-452001, India

**Correspondence:** satishogale@iiserpune.ac.in, suwarnadatar@diat.ac.in

**ABSTRACT**

In view of their immensely intriguing properties, 2D materials are being intensely researched in search of novel phenomena and diverse application interests; however, studies on the realization of 2D/2D nanocomposites in the application-worthy thin film platform are rare. Here we have grown MoS$_2$-hBN 2D/2D composite thin films on different substrates by the pulsed laser deposition (PLD) technique and made comparative studies with the pristine MoS$_2$ and hBN films. The Raman and x-ray photoelectron spectroscopy (XPS) techniques as well as high-resolution transmission electron microscopy (HRTEM) confirm the concomitant presence of both the 1T (conducting) and 2H (semiconducting) polymorphs of MoS$_2$ in the composite film. Interestingly, a peculiar reentrant semiconductor-metal-insulator transition is seen in the MoS$_2$-hBN 2D/2D composite film which is absent in the MoS$_2$ film, and it correlates well with the signatures of phonon softening seen in temperature dependent Raman spectroscopy. Furthermore, electrostatic force microscopy (EFM) reveals the presence of three distinct regions (metallic, semiconducting and insulating) in the MoS$_2$-hBN composite film with differing contact potentials and enhanced propensity for charge transfer with respect to pristine MoS$_2$. A triboelectric nanogenerator (TENG) device containing biphasic MoS$_2$-hBN composite film as an electron acceptor exhibits more than two-fold (six-fold) enhancement in peak-to-peak output voltage as compared to the pristine MoS$_2$ (hBN) film. These observations bring out the potential of 2D/2D nanocomposite thin films for unfolding emergent phenomena and technological applications.

## I. INTRODUCTION

Since the discovery of graphene, there is a significant growth of interest and activity in the field of two-dimensional (2D) materials in view of their unique set of physical properties driven by the surface effects as well as the quantum size effects which render a special nature to their electronic density of states.[1–4]

Among the list of numerous 2D materials of the past and emergent interest, $MoS_2$ is the second most explored material after graphene and has been studied in various applications for over a decade.[5] As an *n*-type transition metal dichalcogenide (TMD) semiconductor, $MoS_2$ exhibits astonishingly diverse, tunable and application-worthy physical, chemical, mechanical and optical properties.[6–13] These include superconductivity, [10] metal-insulator transition (MIT) [11] and topological insulator [12] properties. One of the remarkable feature of $MoS_2$ is that it shows multiple polymorphs, the primary polymorphs being 1T (trigonal phase), 1H and 2H (hexagonal phase), and 3R (rhombohedral phase). Among these, the 1T polymorph, which is a paramagnetic metal[14], is thermodynamically the least stable one and tends to form the stable 2H phase by reorganizing its stacking layers. Ubiquitously, coexistence of two or more structural phases can influence the material functionalities in interesting ways.[15–18] However, stabilizing the multiphase state (e.g. 1T and 2H phases of $MoS_2$) is challenging due to the differing stabilizing conditions for different phases. Various synthetic methods e.g. chemical exfoliation, hydrothermal method and chemical vapour deposition (CVD) have been examined to achieve the multiphase states.[17,18] Importantly, it has been shown that $MoS_2$ exhibit phase change from semiconductor (2H) to metal (1T), enhanced hydrogen evolution reactions (HER) activity due to significantly higher electrical conductivity, useful for device fabrications.[19,20]

Although several interesting works have been reported in the literature on the growth of large area thin films of 2D materials and their heterostructures,[7] attempts at the growth of multiphase stabilized 2D/2D nanocomposites in the device application-worthy *thin film* form of contemporary device interest are rare [21]. There is thus a quest for a new route to realize this objective. It is also of great interest to explore the application potential of the films which could be more in the form of mosaics of nanoscale sheets (turbostratic configuration), rather than a single or few layer large area uniform coatings. Such a configuration renders a natural benefit of retaining and expressing the properties of *few layer forms* of these 2D materials in a thin film assembly. In this context, the *pulsed laser deposition* (PLD) technique allows a great flexibility of stoichiometry, layer thickness, and growth temperature/ambient control, enabling growth of different thin film configurations, their hetero-structures or even super-lattices. Keeping this in mind, we have chosen hexagonal boron nitride (hBN); another prominent layered 2D material to grow nanocomposite films of $MoS_2$ and hBN, with a hope to induce and stabilize multiphase in $MoS_2$ (1T and 2H phases in the present study and henceforth we will call it as *biphasic*) because of the specific PLD growth dynamics as well as strong interlayer coupling between $MoS_2$ and hBN.[22] hBN is also intensely researched material due to its exotic optoelectronic properties, mechanical robustness and thermal stability. It has been extensively studied for various device applications, e.g. field effect transistor, detectors and photoelectric devices.[22,23] It is an insulator with a large band gap of ~6 eV, has atomically flat surface without any dangling bonds, charge impurities, and importantly it is chemically inert.[24–26]

In this article, relying on the specific strengths and advantages of the PLD technique, we have successfully stabilized biphasic 2D material by growing thin films of 2D/2D $MoS_2$-hBN system using a single composite target. Biphasic nature was confirmed by Raman and x-ray photoelectron spectroscopy. Temperature dependent Raman spectra show the signature of phonon mode softening with decreasing temperature. Interestingly, in our $MoS_2$-hBN composite case, the temperature dependent resistivity shows an intriguing reentrant semiconductor-metal-insulator transition as a consequence of competing polymorphs, which is absent in $MoS_2$ film. Electrostatic force microscopy (EFM) reveals the presence of three distinct regions (metallic, semiconducting and insulating) in the composite films with the observation of three different contact potentials, respectively. In addition, EFM study without and with charge injection on the surface reveals the electron acceptor character of the composite case with enhanced propensity for charge transfer vis a vis the pristine $MoS_2$. Furthermore, a triboelectric nanogenerator (TENG) device containing biphasic $MoS_2$-hBN thin film as an electron acceptor shows more than two-fold (six-fold) increase in the output voltage with respect to the pristine $MoS_2$ (hBN) thin films.

## II. THIN FILM GROWTH

For this study, we have grown ~200 nm pristine $MoS_2$, hBN and composite $MoS_2$-hBN thin films on $c$-$Al_2O_3$ (0001) and n-Si substrates by PLD (KrF laser, wavelength = 248 nm). Schematic of a typical PLD growth is shown in **Fig. 1**. A polycrystalline target of $MoS_2$-hBN was made by using the solid state reaction method. For this, $MoS_2$ and hBN powders were mixed (1:1 molar ratio) and sintered in Argon atmosphere for 8 hours at 750 °C. $MoS_2$ and hBN targets were also made by following the conventional solid state reaction methods. Prior to the deposition, the substrates were annealed at 1000 °C for one hour to make them atomically flat as confirmed by atomic force microscopy (AFM) revealing clear step features (see supplementary **Fig. S1**). The target to substrate distance was kept ~ 40 mm. All the films were grown at a substrate temperature of 500 °C at a pulse repetition rate of 10 Hz using laser energy density of 0.5 J/cm$^2$, in the presence of 5 mTorr Argon pressure. Additionally we have also grown films on flexible Kapton substrates (grown at 400 °C while keeping the other growth conditions same) for device fabrications.

In the PLD process depicted in **Fig. 1**, it should be noted that the pulses of 5 eV photons dump significant amount of energy (density) in 20 ns time scale onto the substrate surface, causing highly non-equilibrium phenomena leading to the formation of a plasma plume wherein various ionic and radical species are present. They progress towards the hot substrate with a fairly high velocity distribution. On the substrate surface, the impinging radicals seek equilibrium or near-equilibrium phase formation based on the temperature and simultaneously impinging atoms/ions of the ambient. The semi-equilibrium solubility and miscibility criteria of the potentially developing phases then governs the final phase fractions and their character in the film. As discussed and evidenced through detailed characterizations later, this process leads to a state wherein N doped $MoS_2$ sheets coexist with hBN sheets in the form of a mosaic or turbostratic assembly. It is interesting to note that although Mo, S, B, N related ions and radicals impinge on the surface during PLD, the growth equilibria render the two chemically distinct separate phases.

## III. RESULTS AND DISCUSSION

### A. Raman spectroscopy

First, the Raman spectra of $MoS_2$-hBN thin films were recorded with a 2.33 eV (532 nm) excitation energy laser (**Fig. 2**). The main $E_{1g}$, $E_{2g}$ and $A_{1g}$ peaks were observed at 285.4, 381.9 and 408 cm$^{-1}$, which correspond to the vibration of atoms along in-plane ($E_{1g}$, $E_{2g}$) and out-of-plane ($A_{1g}$) directions, respectively. It is clearly seen that the $E_{2g}$ peak is red-shifted by ~1.5 cm$^{-1}$ as compared to the pristine case and splits into two peaks, labeled as $E_{2g}^{1+}$ and $E_{2g}^{1-}$. This peak splitting is attributed to the symmetry breaking of $E_{2g}$ vibrational mode as a consequence of the introduction of lattice strain due to in-situ N doping and interface formation with hBN.[27] We also observed significant amount of red shift (~4 cm$^{-1}$) in the $A_{1g}$ mode. The shifts in the $A_{1g}$ and $E_{2g}$ modes in biphasic $MoS_2$-hBN thin film with respect to the pristine $MoS_2$ reflect the sensitivity of these modes towards to strain and doping in $MoS_2$.[28, 29] The shifts and broadening also indicate strong interlayer coupling between $MoS_2$ and hBN. More importantly, the observed significant enhancement of the $J_1$, $J_2$ and $J_3$ phonon modes at lower wave numbers in the composite film (**Fig. 2** and **Fig. S2**) as compared to the pristine $MoS_2$ case implies significant 1T component in the composite case.[30] We also observed that the characteristic peak of the $E_{2g}$ phonon mode at ~1365 cm$^{-1}$ in the pristine hBN case is red shifted by ~2 cm$^{-1}$ in the composite case, and appears significantly weaker and broader implying the random strain generated during the growth.[31]

The Raman spectroscopy data for the $MoS_2$ and $MoS_2$-hBN at selected temperatures (200 K ≤ T ≤ 300 K) are shown in **Fig. 3(a)** and (**b**). These data are later discussed in the context of the interesting reentrant resistivity transition. The modes were fitted using Lorentzian function and the position of the $A_{1g}$ mode thus obtained is plotted as a function of temperature in **Fig. 3(c)**. The mode position of the $MoS_2$ shows a linear behavior apart from small but noticeable deviation around T = 250 K. The temperature coefficient of the frequency of this mode obtained by applying a linear fit to the data points results in the rate coefficient ($\chi$) as -0.019 cm$^{-1}$K$^{-1}$. This value is close to the one reported for $MoS_2$.[32] In **Fig. 3(d)**, the wavenumbers corresponding to the $A_{1g}$ mode of the $MoS_2$ and $MoS_2$-hBN layers are plotted as a function of temperature. It is seen that the modes show a red shift with increasing temperature for the $MoS_2$ layer, as reported before. Such a behavior is attributed to the anharmonic contributions to the interatomic potential energy, mediated by phonon-phonon interactions. On the other hand the position of the $A_{1g}$ mode for the MoS$_2$-hBN layer shows a very subtle reduction when temperature is increased. Interestingly, the mode position shows a significant reduction above $T = 240$ K. The first principle density functional theory (DFT) calculations on MoS$_2$ have clearly established that electron-phonon coupling is stronger in the case of $A_{1g}$ mode than the $E_{2g}$ mode.[29] In another *ab initio* study of the interatomic force constants, separating the short-range and the long-range Coulomb parts, it is shown that the dielectric screening decreases the frequency of optical phonon modes.[33] Hence it can be expected that an increase in dielectric constant and dielectric screening would occur, as the system contains insulating hBN layers. This increased dielectric screening is thus causing the decrease in frequency of the Raman mode below $T = 240$ K. Thus, our temperature dependent Raman spectroscopy studies clearly show a modulation in phonon frequency of the $A_{1g}$ mode. Following Chakraborty *et al.*, [29] the $A_{1g}$ mode is highly sensitive to the electron-phonon interaction which is naturally going to affect the electronic transport due to softening of the mode frequency.

## B. X-ray photoelectron spectroscopy

The XPS data for the individual cases of the MoS$_2$ and hBN films along with that for the composite film are shown in **Fig. 4(a)**. Specifically, the Mo 3p characteristic in MoS$_2$ films solely represents the presence of the 2H phase. The XPS data of **Fig. 4(b)** for the composite film (dark red) shows broader distribution of the primary Mo 3p$_{3/2}$ contributions with some structure. In the same figure we have shown the directly (arithmetically) added (MoS$_2$ + hBN) contributions from **Fig. 4(a)**, to reveal departure from the simply mixed spectral states due to the composite formation. It is clear that the departure emanates primarily from the development of complex Mo-N coordination reflecting doping of N in MoS$_2$.[27] Another possibility of the presence of Mo (IV) via MoS$_3$ coordination cannot be ruled out.

In the Mo 3*d* region of the data for the composite case, the contributions again appear somewhat broader and can be fitted by multiple peaks as shown in **Fig. 4(c)**. Interestingly, the two intense peaks at 229.6 eV (Mo 3d$_{5/2}$) and 232.6 eV (Mo 3d$_{3/2}$) correspond to the 2H phase of MoS$_2$ with Mo$^{4+}$ state, [17, 24] while the two other peaks (@ 228.5 and 231.5 eV), relatively shifted to lower binding energies by ~1 eV, reflect the existence of 1T phase.[34, 35] This unambiguously confirms the coexistence of 2H (semiconducting) and 1T (metallic) phases in the composite sample. From the area coverage of the peaks, the phase proportion can be estimated to be 58% (1T): 42% 2H phase. In comparison, the Mo 3*d* region of the data for the pristine MoS$_2$ case clearly depicts the 2H phase of MoS$_2$ film (**Fig. S3**).

The S 2p core data for the composite case is presented in **Fig. 4(d)**. Two broad doublets are seen in the spectrum which can be fitted by four peaks. The two doublets with the binding energies of 161.7 and 163.4 eV reveal the presence of 2H phase of MoS$_2$ while the other two peaks at 160.8 eV and 162.6 eV reflect the presence of metallic 1T phase. Thus, the biphasic nature of the composite is further confirmed from the S 2p spectral analysis as well.

## C. High-resolution transmission electron microscopy, field emission scanning electron microscopy and X-ray diffraction

To elucidate the microscopic constitution of the films, we employed field emission scanning electron microscopy (FESEM) with EDAX and high resolution transmission electron microscopy (HRTEM) measurements. **Fig. 5a** along with its inset present the FESEM data (top and side views) for the MoS$_2$-hBN film grown on sapphire substrate (typical grain size 50 nm), while the corresponding data for MoS$_2$ is shown in **Fig S4.** These data establish that the films are quite dense in all the cases, although the surfaces are noted to exhibit overgrowth-like features in the case of MoS$_2$-hBN composite film and flaky features in the case of only MoS$_2$ film. Due to significant charging during microscopy measurements, the clean images could not be obtained in the case of h-BN films. Compositional analysis though EDAX is shown in **Fig. 5b**. Technically, through EDAX, it is difficult to obtain the reliable compositional ratio of B and N as these are low-Z elements.

In view of the non-availability of cross-section TEM imaging capability, for recording HRTEM data an interesting deposition strategy was adopted wherein a carbon coated holy grid was directly used as a substrate and an ultrathin film of the composite was grown thereupon. The HRTEM images of **Fig. 5c** and **Fig. 5d** clearly show that ultrathin sheet-like features are seen to have assembled during the PLD growth, which confirms the mosaic or turbostratic arrangement. The lattice image and its analysis shown in **Fig. 5e** further shows that the film is indeed in the form of a 2D/2D heterostructure of MoS$_2$ and h-BN. Confirmation of the coexistence of the 1T and 2H phases in the sample was also obtained from the HRTEM analyses. Indeed, the biphasic nature of MoS$_2$ could be clearly revealed by the selected area electron diffraction (SAED) pattern (**Fig. 5f**), with the spots identified with red line representing the 2H-MoS$_2$, whereas the spots identified with the yellow line representing the 1T phase. The 2H phase has trigonal prismatic arrangement of Mo and S with AbA and BaB stacking, whereas the 1T phase has octahedral arrangement with AbC and AbC stacking. Therefore, in the 1T case the diffraction spots are usually angularly shifted from there ideal place in the basal plane by 30° (yellow line).[36] As stated by Reshmi *et al*.,[37] the *d*-spacing of 2.34 Å (**Fig. 5e**) closely corresponds to (2 0 1) and (1 0 3) lattice constants of 1T and 2H phase(s) of MoS$_2$, albeit with a degree of strain. The HRTEM of pristine cases MoS$_2$ and hBN shown in **Fig. S5** also shows the corresponding *d* value of 2.2 Å and 2.5 Å respectively.

As stated above, the HRTEM images of **Fig. 5c** and **Fig. 5d** show that the growth appears to be in the form of a dense mosaic of ultrathin MoS$_2$ and hBN sheets. This implies that the orientation of the film should be reflected to be along the c-axis in the XRD pattern. This is indeed the case, as shown in **Fig. 5g** and for the pristine MoS$_2$ case (**Fig. S6**).[38]

**D. Electronic transport properties**

Electronic transport in 2D material is known to be very sensitive to the external perturbations (e.g. pressure, strain, liquid gating).[6–9] For example, 2H phase MoS$_2$ shows a semiconductor to metal transition as a consequence of reduction in S-S interlayer distance, induced by the applied high external pressure of ~19 GPa.[39] Since our composite film consists of both the semiconducting (2H) and metallic (1T) phases as well as doping of nitrogen at the sulphur site (as seen from XPS and Raman data), it is highly interesting to investigate the electrical transport of the composite film.

Indeed, the result of the temperature dependent in-plane four probe resistivity measurement of the biphasic MoS$_2$-hBN thin film (**Fig. 6a**) shows a highly peculiar semiconductor-metal-insulator re-entrant transition in resistivity. This can certainly be attributed to the presence of metallic 1T phase that gets stabilized in the composite case alone, while the MoS$_2$ itself leads only to the semiconducting 2H phase and shows the semiconductor-insulator transition while lowering the temperature. While cooling the biphasic film from $T = 300$ K, it shows a semiconducting behaviour, while within the temperature range of 225 K $\leq T \leq$ 250 K, a drop is observed in resistivity, a clear signature of a semiconductor-to-metal transition. Below $T \leq 225$ K, again it makes a transition to the insulating state. Similar feature in the electronic transport was also observed by applying very high external pressure upto ~35 GPa.[40] In the metallic region, the resistivity follows linear temperature dependent ($\rho \propto T$) (**Fig. S7a**) as generally found in a normal metal. The insulating region was fitted with the 3D variable range-hopping model ($\ln\rho \propto T^{-1/4}$) (**Fig. S7b**) associated with the presence of disorder. The activation energy ($E_A$) was found to be ~0.145 eV. Similar behaviour was also observed in pristine $MoS_2$ under very high pressure as a consequence of competing phases as well as occurrence of charge density waves (CDW).[41]

We also examined the current-perpendicular-to-plane (CPP) transport in the $MoS_2$ and $MoS_2$-hBN films in view of the intrinsic anisotropic nature of the films. The CPP transport was measured across $MoS_2$/n-Si and $MoS_2$-hBN/n-Si interface by sweeping voltage from -1V to 1V at scan rate of 62 mV/s. In the CPP transport, the top contact for the I-V measurements was gold (Au). The corresponding data with n-Si bottom and Au contact on the top is shown in **Fig. 6b**, with the convention that the voltage is considered positive (on the *x*-axis) when a positive potential is applied to the top of composite film. It is immediately noted that the current at a given voltage is significantly (over an order of magnitude) higher under reverse bias and also substantially higher under forward bias. This might be attributed with the fact that conductivity contribution emanating from the metallic 1T phase present in the biphasic sample is not laterally connected to enable percolation, but in the vertical configuration, it could either connect/percolate or contribute via tunnelling through ultrathin hBN sheets in the layered mosaic composite films. The bias-asymmetry of the curves is seen to be reversed between the $MoS_2$ and $MoS_2$-hBN composite cases. Given the same contacts on the two sides in both the cases, this feature suggests a change in the nature of majority carriers. Interestingly, the resistivity difference is not noted to be as significant in the case of four probe in-plane resistivity shown in **Fig. 6a**. This asymmetry between the in-plane and CPP transport signifies the presence of a specific constitution of the conducting 1T phase component.

**E. Electrostatic force microscopy**

Since a composite film comprising of mixed semiconducting and metallic polymorphs of $MoS_2$ and insulating hBN layers, has multiple interesting interfaces, these can have interesting consequences for the charge distributions in the layer and the corresponding response to external applied field. To confirm the distribution and field response of the electronic charges over the surface, we performed the electrostatic force microscopy (EFM) analysis. EFM was performed in two-pass lift off mode. First pass was for AFM (atomic force microscopy) morphology study and second pass for EFM was at the height of 50 nm. Qualitatively the topography image shows surface roughness of the sample and EFM phase image shows the electrical nature of the sample in different regions. Different phases present in the EFM data depict surface potential or work function difference between various components in the sample. The topography and the corresponding EFM phase image of composite sample at bias voltage of 3V were also obtained (**Fig. S8 (a)** and **(b)**). One can identify at least three different phase contrasts (A, B and C) in the image marked by circles; also confirmed from the histogram showing three maxima (**Fig. S8(c)**). Quantitatively EFM phase shift is directly related to the local surface potential described as

$$\Delta\phi = -\arcsin\left[\frac{Q}{2k}\frac{d^2C}{dz^2}(V_{tip}-V_{surface})^2\right]$$

Where $\Delta\phi$ is the phase shift, $Q$ quality factor and $k$ is the force constant of the cantilever, $\frac{d^2C}{dz^2}$ is the second derivative of the capacitance between the sample and the probe as a function of the vertical distance, $V_{tip}$ is the DC bias applied to the tip, and $V_{surface}$ is the local surface potential on the sample surface.[42] To estimate the local surface potential variation in hBN, $MoS_2$ and $MoS_2$-hBN, a plot of phase shifts at different biases between the film and the tip, measured by EFM-phase, with a fixed lift height of 50 nm was obtained and fitted with the least square measurement to the above mentioned equation as shown in the **Fig. 7(a)** and **(b)**. All the curves are parabolic which confirms the proportional interaction variation with $(V_{tip}-V_{surface})^2$ which is indicative of contact potential difference (CPD) between tip and the surface. Surface potential can be attributed to charge distribution in real space or work function (WF) difference between the tip and the sample depending on whether the sample is insulating, semiconducting or conducting. Several factors can contribute to the local charge distribution in the case of insulators, like humidity, tip-sample interaction, adsorption of charged ions etc. As indicated in **Fig. 7(c)**, hBN films have surface potential of -0.46 eV, which could be because of adsorbed charges from the ambient condition. Therefore, to better understand the nature of the charges, surface potential of hBN is plotted with respect to the applied bias voltage to the tip which clearly shows that change in surface potential after negative polarization is much greater than that of the positive polarization, indicative of presence of trapped sites for negative charges present in the sample. Further, as shown in the same figure, the contact potential difference for $MoS_2$ is 1.3V. Assuming the tip work function to be 5.6 eV for platinum tip, the work function of $MoS_2$ films can be estimated to be 4.3 eV which matches with the reported work function for $MoS_2$ in ambient conditions. Further, CPD for the composite was observed to be lowered as shown in **Fig. 7(c)**. To investigate this further, data was taken at various sites on the composite which indicated different phases in the image. The fitted data for the composite as shown in figure shows that work function of the $MoS_2$ seems to have increased; showing at least two distinct work function values of 4.87 eV and 5.09 eV (**Fig. 7(c)** and **(d)**). It indicates that two different phases of $MoS_2$ have indeed been formed, as also observed other techniques.

To further investigate the effect of hBN in biphasic $MoS_2$-hBN film on the formation of constituents with different work functions in the film, charges were injected in the film using a conducting tip, following a procedure discussed by Zhao *et al*.[43] This was done by taking the tip towards the sample in contact mode with applied bias voltage of ± 10V and then later imaging the area for smaller applied bias voltage. It was observed that there was no change in the EFM image after injecting charges by applying +10V, whereas one can clearly see that charges have been injected in the film for -10V clearly indicating that there are trapped sites for negative charges which have been introduced after the formation of composite (**Fig. 8**). The topographic image and corresponding line profile, before and after charging is shown in **Fig. S9**. However, no influence of the charge injection was found on pristine $MoS_2$ film (**Fig. S10).** Therefore, hBN and related interfaces seem to have provided sites for trapped charges by means of formation of interfaces of different work functions which are related with the changes in CPD, affecting the electron transfer process. As extensively discussed by Seol *et al*.,[20] such distributions can have interested consequences for triboelectric nanogenerator (TENG) applications of 2D materials. Hence, we proceeded to examine this aspect in a preliminary study, as discussed below.

 F. Triboelctric nanogenerator

Triboelectric nanogenerators (TENG) devices based on 2D materials is becoming a cutting-edge research area for future mechanical energy harvesting. Therefore, based on the EFM data discussed above, we explored our biphasic $MoS_2$-hBN thin films as an electron acceptor in a TENG configuration in comparison with the pristine $MoS_2$ case.[20] We designed a TENG device as shown in **Fig. 9(a)**. For this we grew biphasic $MoS_2$-hBN (and also $MoS_2$ and hBN) films on a flexible Kapton substrate at 400 °C, by keeping all the other growth conditions same in PLD. The lower growth temperature for Kapton substrate as compared to the 500°C used for sapphire or silicon substrate was based on the consideration of thermal stability of Kapton. The morphology of the film was found to be slightly rougher as compared to that on the sapphire and silicon substrates, however the Raman spectrum (**Fig. S11**) showed that the basic character of the film was essentially of biphasic nature. The small difference noted was the lower tailing of the $E_{2g}$ peak suggested to be due to nitrogen doping of $MoS_2$. Interestingly the other signatures of the 1T phase (namely $J_1$, $J_2$ and $J_3$) are quite well defined in this case.

The $MoS_2$-hBN on kapton and polypropylene sheet on top of it was sandwiched between two aluminum layers; copper wires were soldered to the electrodes from the top and bottom sides. The area of the sample was 5 mm x 10 mm. The mechanical input was applied by a vibrator @ 50 Hz with a measured force of 10 N (**Fig. 9b**). The output open circuit voltage (no load resistance) values obtained for hBN, $MoS_2$ and $MoS_2$-hBN are shown in **Fig. 9c**; while the same is shown on the expanded scale for clarity about the shape of the signals. It can be clearly seen that the open circuit voltages for hBN, $MoS_2$ and $MoS_2$-hBN are in the ratio of 6: 3: 1. Thus, there is a two-fold increase of average peak-to-peak output voltage for the $MoS_2$-hBN (~14.7 V) case *vis a vis* pristine $MoS_2$ (~7.2 V). In contrast, composite film shows six-fold increase in of average peak-to-peak output voltage with respect to pristine hBN (~2.3 V). This is consistent with the propensity of biphasic film for charging effects under field as reflected by the EFM data. We also estimated the powering capability of the devices based on the three cases by adding a load of 1 MΩ (**Fig. S12**). Here again the composite case shows a higher power delivery. A comparison table in terms of overall TENG performance based on 2D materials is as given in **Table. S1**. Finally but interestingly, the shape of the TENG voltage signal curves is peculiar and has been noted before only in the case of 2D materials.[44] We will undertake further studies to optimize and enhance the TENG performance in subsequent studies.

**IV. CONCLUSIONS**

Summarizing, we have used the pulsed laser deposition (PLD) technique to stabilize the metastable metallic 1T phase in significant proportion (>50%) along with the highly stable 2H phase of $MoS_2$ in supported thin film grown on single crystalline substrates. Our detailed chemical and microstructural characterizations by XPS, Raman spectroscopy and HRTEM confirm the presence of biphasic nature of the composite films. Temperature dependent Raman spectra show the signature of phonon mode softening and bear interesting correlation with the transport property, exhibiting a peculiar re-reentrant semiconductor-metal-insulator transition in the temperature dependent resistivity only in the case of composite film. Electrostatic force microscopy without and with charge injection reveals the presence of three distinct regions (metallic, semiconducting and insulating) in the composite film with differing contact potentials and electron acceptor character. A triboelectric nanogenerator device containing biphasic $MoS_2$-hBN composite film as an electron acceptor exhibits more than two-fold (six-fold) enhanced peak-to-peak output voltage as compared to the pristine $MoS_2$ (hBN). This study will pave the way to engineer new multiphasic 2D heterostructures via thin film growth to obtain novel multi-functionalities.

ACKNOWLEDGMENTS

Swati Parmar would like to thank CSIR for SRF fellowship. Satish Ogale would like to thank DST Nanomission Thematic unit for funding support and the Department of Atomic Energy for the award of Raja Ramanna Fellowship and grant. Abhijit Biswas would like to thank the government of India for providing the SERB-N-PDF fellowship (PDF/2017/000313). We would like to thank Prof. Surjeet Singh and Ms. Prachi Telang for the R-T measurement. We would also like to thank Mr. Sharad Varma for the XPS measurements and Mr. Anil Shetty for FESEM measurement.

**FIGURE CAPTIONS**

**FIG. 1.** (Color-online-only) Schematic of the pulsed laser deposition set up for growing $MoS_2$, hBN and $MoS_2$-hBN thin films. Plasma generated after laser ablation of target consists of high energy (0.1-10 eV) ions and radicals of Mo, S, B and N.

**FIG. 2.** (Color-online-only) Raman spectra of $MoS_2$, hBN, and $MoS_2$-hBN composite thin film grown on $c$-$Al_2O_3$ substrate, showing the presence of both 1T and 2H phases of $MoS_2$. Left side of the $x$-axis break is for $MoS_2$ signatures (100 to 500 $cm^{-1}$) and the right side of the break is for hBN (1300 to 1500 $cm^{-1}$).

**FIG. 3.** (Color-online-only) (a)-(b) Temperature dependent Raman spectra of $MoS_2$ and $MoS_2$-hBN thin films, respectively, within the temperature range of 200 K $\leq T \leq$ 300 K; (c) Temperature dependence of $A_{1g}$ phonon mode of $MoS_2$; (d) Comparison between temperature dependence of $A_{1g}$ phonon mode of $MoS_2$ and biphasic $MoS_2$-hBN composite film.

**FIG. 4.** (Color-online-only) XPS spectra of $MoS_2$, hBN and $MoS_2$-hBN composite thin films grown on $c$-$Al_2O_3$ substrates; (a) N-1s and Mo 3p core of $MoS_2$, hBN and $MoS_2$-hBN composite films; (b) N1s and Mo 3p core of arithmetically added ($MoS_2$ + hBN) contribution compared with $MoS_2$-hBN composite case; (c) Mo+S 3d core of $MoS_2$-hBN; and (d) S-2p for $MoS_2$-hBN.

**FIG. 5.** (Color-online-only) (a) Top and side view (inset) FESEM images of $MoS_2$-hBN film grown on $c$-$Al_2O_3$; Typical grain size is ~50 nm (inset) (b) EDAX analysis of $MoS_2$-hBN grown on $c$-$Al_2O_3$ film confirms 2:1 compositional ratio of Mo and S in the film, as well-known the lighter (low-Z) elements (B, N) stoichiometry is not truthfully revealed by EDAX; (c)-(e) HRTEM images of $MoS_2$–hBN deposited on carbon coated TEM grid at room temperature; the $d$ spacing were found to be 2.34 Å (for $MoS_2$) and 2.5 Å (for hBN); (f) Selected Area Electron Diffraction

(SAED) patterns with 30° angle-rotated spots confirm the concurrent presence of 1T and 2H phase of $MoS_2$; (g) X-ray diffraction shows the oriented nature of $MoS_2$-hBN grown on $c$-$Al_2O_3$ film.

**FIG. 6.** (Color-online-only) (a) Resistivity of $MoS_2$ and biphasic $MoS_2$-hBN composite thin films grown on c-$Al_2O_3$ substrate; (b) Current perpendicular to plane (CPP) transport of $MoS_2$ and $MoS_2$-hBN thin films grown on n-type Si (001) substrate. The hBN films are highly resistive, hence the corresponding data not shown.

**FIG. 7.** (Color-online-only) (a) Electrostatic force microscopy of Phase vs. Voltage spectra calculated from EFM images of the surface of $MoS_2$, hBN and biphasic $MoS_2$-hBN film grown on $c$-$Al_2O_3$ for the applied voltages (-5 to +5V), depicting the clear difference in work function compared to prestine cases; (b) EFM phase shift of $MoS_2$-hBN for the applied voltages (-5 to +5V) shows the asymmetric nature as well as the minor difference in the work function indicative of the charge distribution in the film at different areas. The solid parabolic curve are the least square fit to the data; (c)-(d) Contact potential difference in three different films and in three different regions of the composite $MoS_2$-hBN film.

**FIG. 8.** (Color-online-only) EFM images of $MoS_2$-hBN film at tip bias 4V: (a) topography image; (b) EFM phase image before charging; (c) EFM phase image after charging at -10V.

**FIG. 9.** (Color-online-only) (a) Schematic of the device architecture of TENG made of composite biphasic $MoS_2$-hBN composite thin films; (b) TENG performance measurement set up; (c) The continuous output voltage signal of TENG devices ($MoS_2$, hBN and $MoS_2$-hBN) recorded at 50 Hz frequency.

**FIGURES**

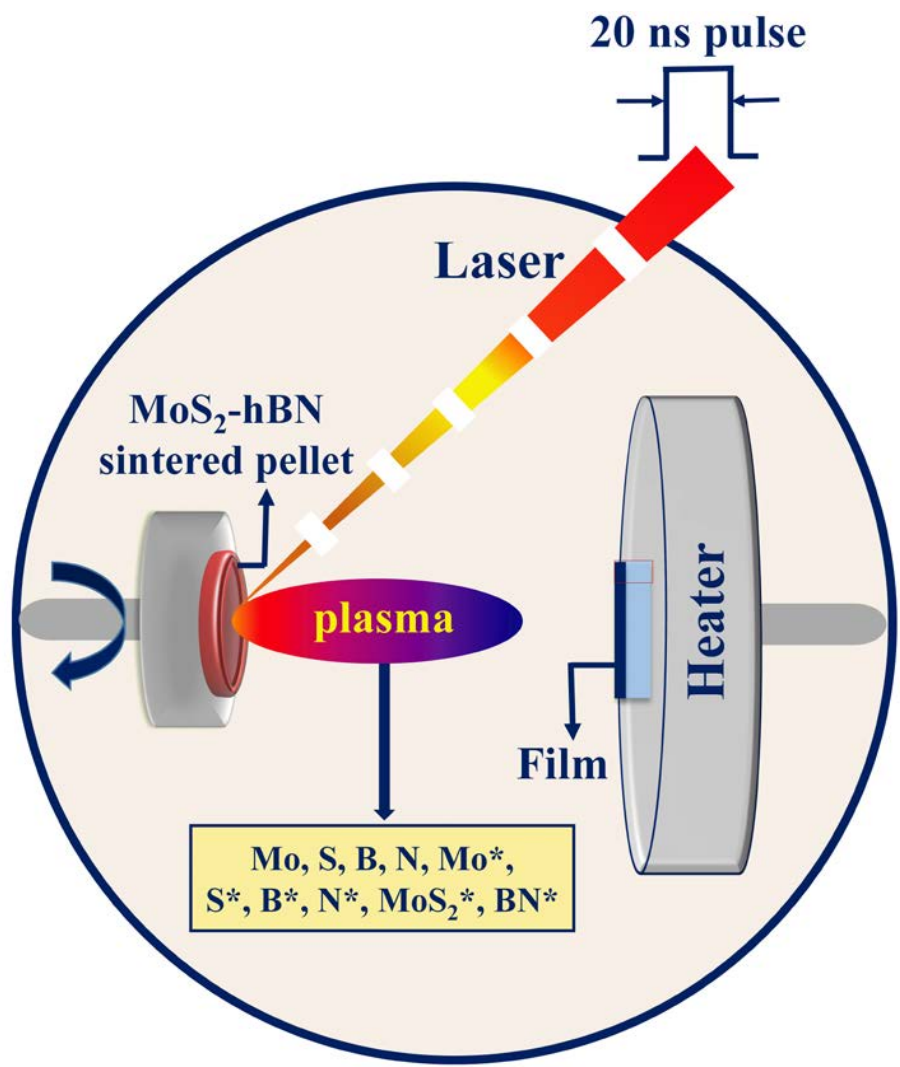

**(Figure-1)**

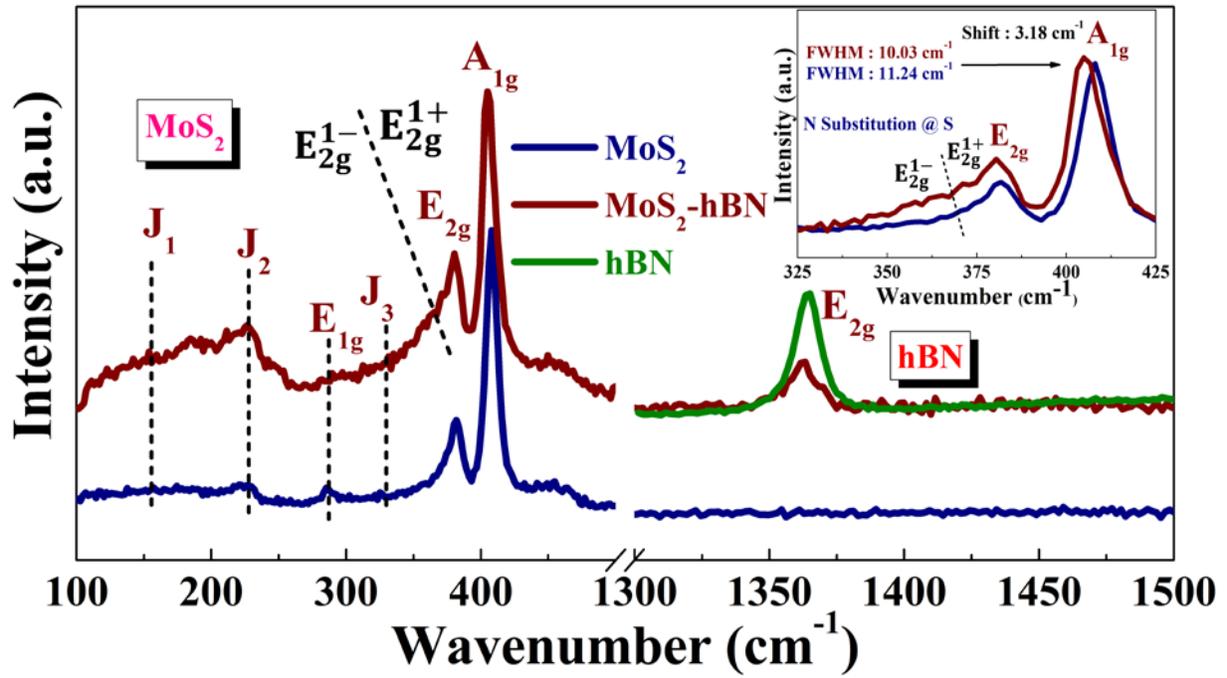

**(Figure-2)**

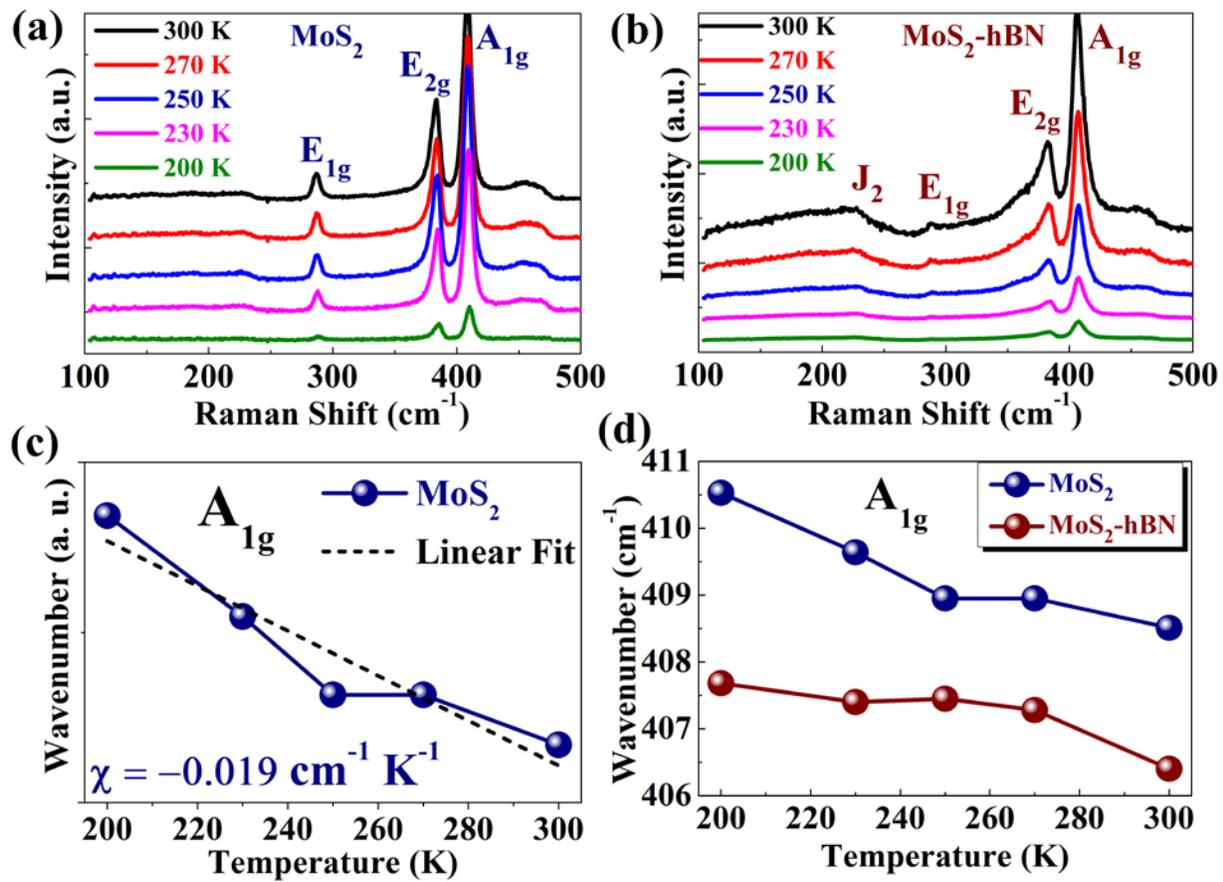

**(Figure-3)**

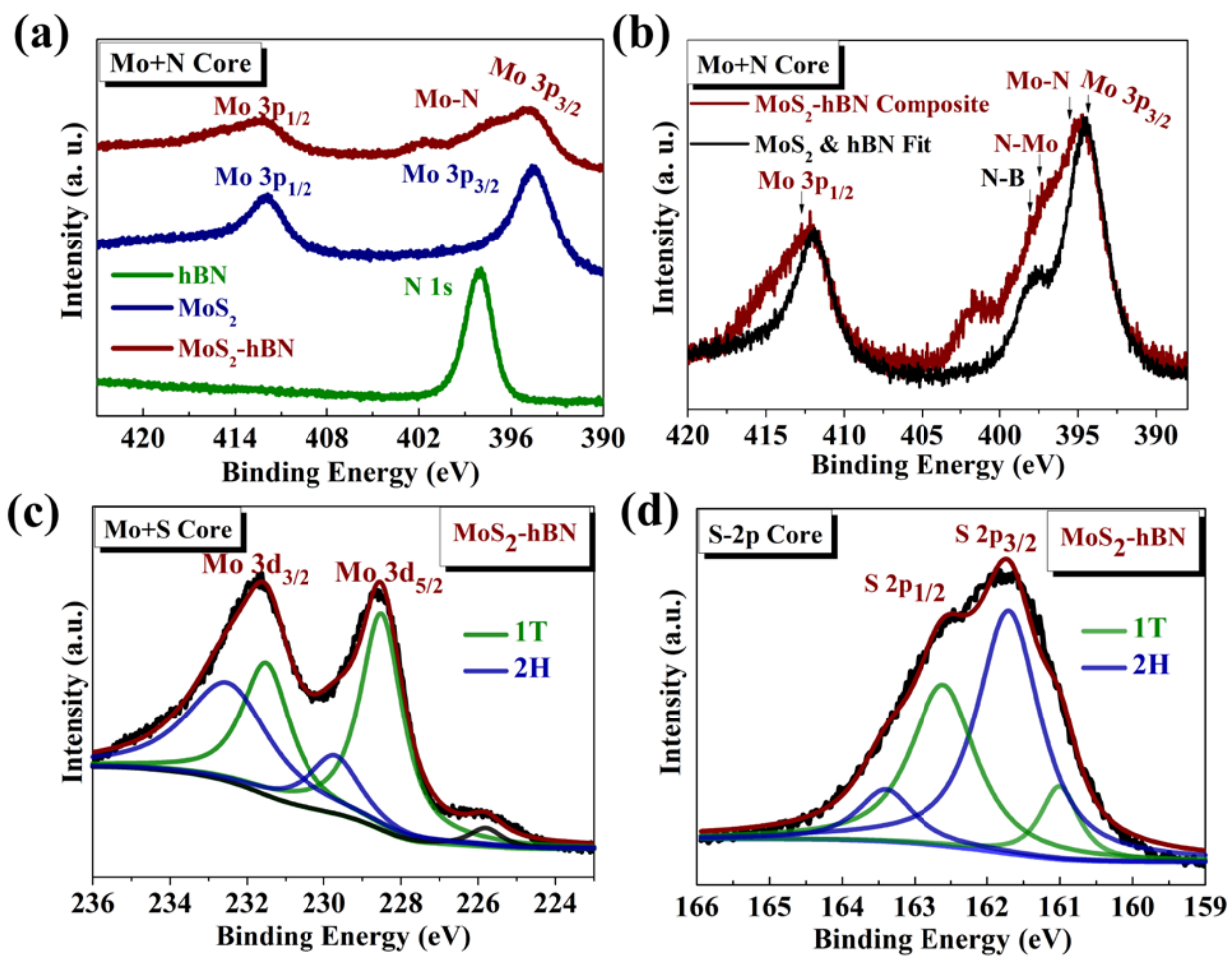

**(Figure-4)**

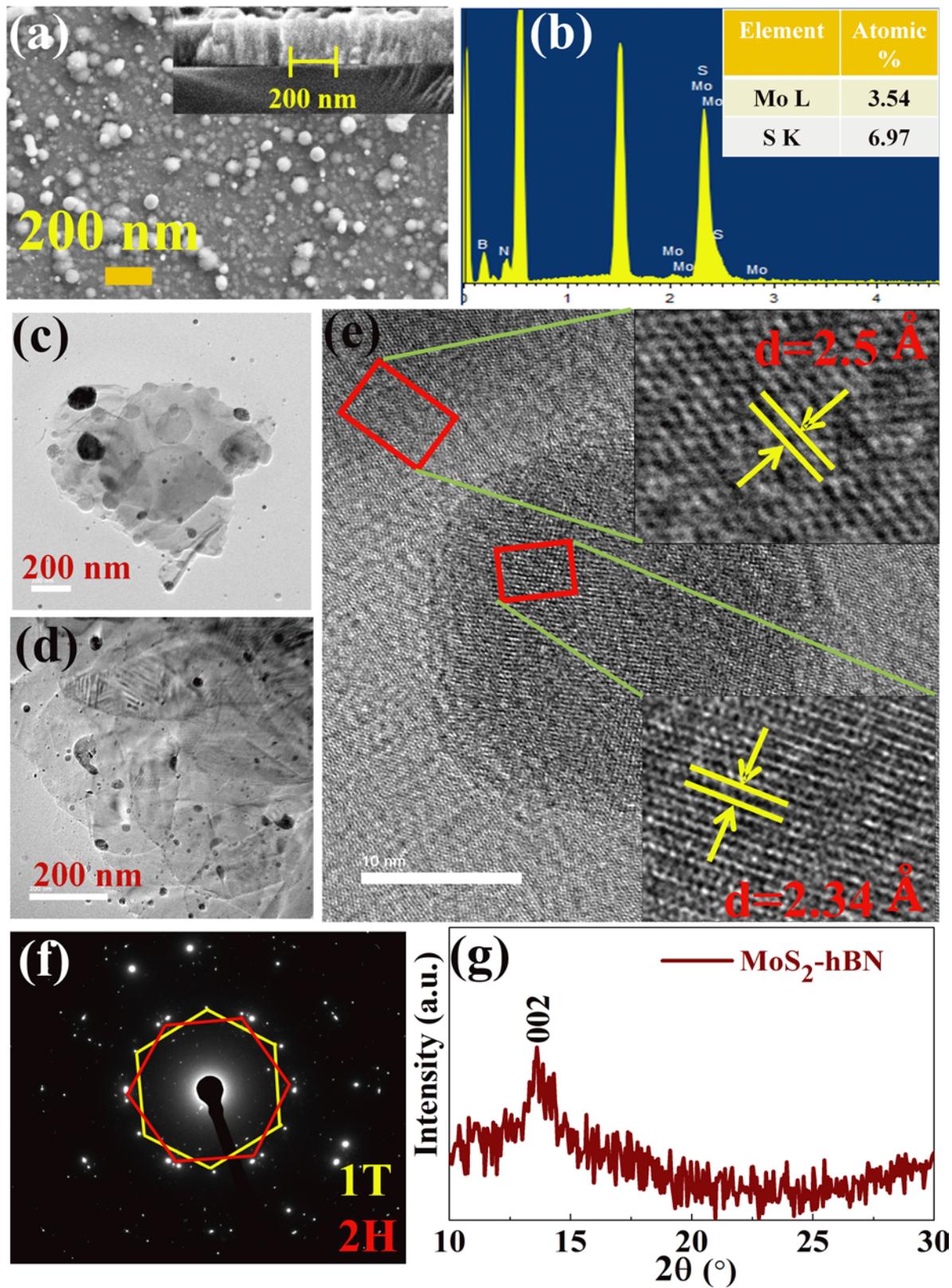

**(Figure-5)**

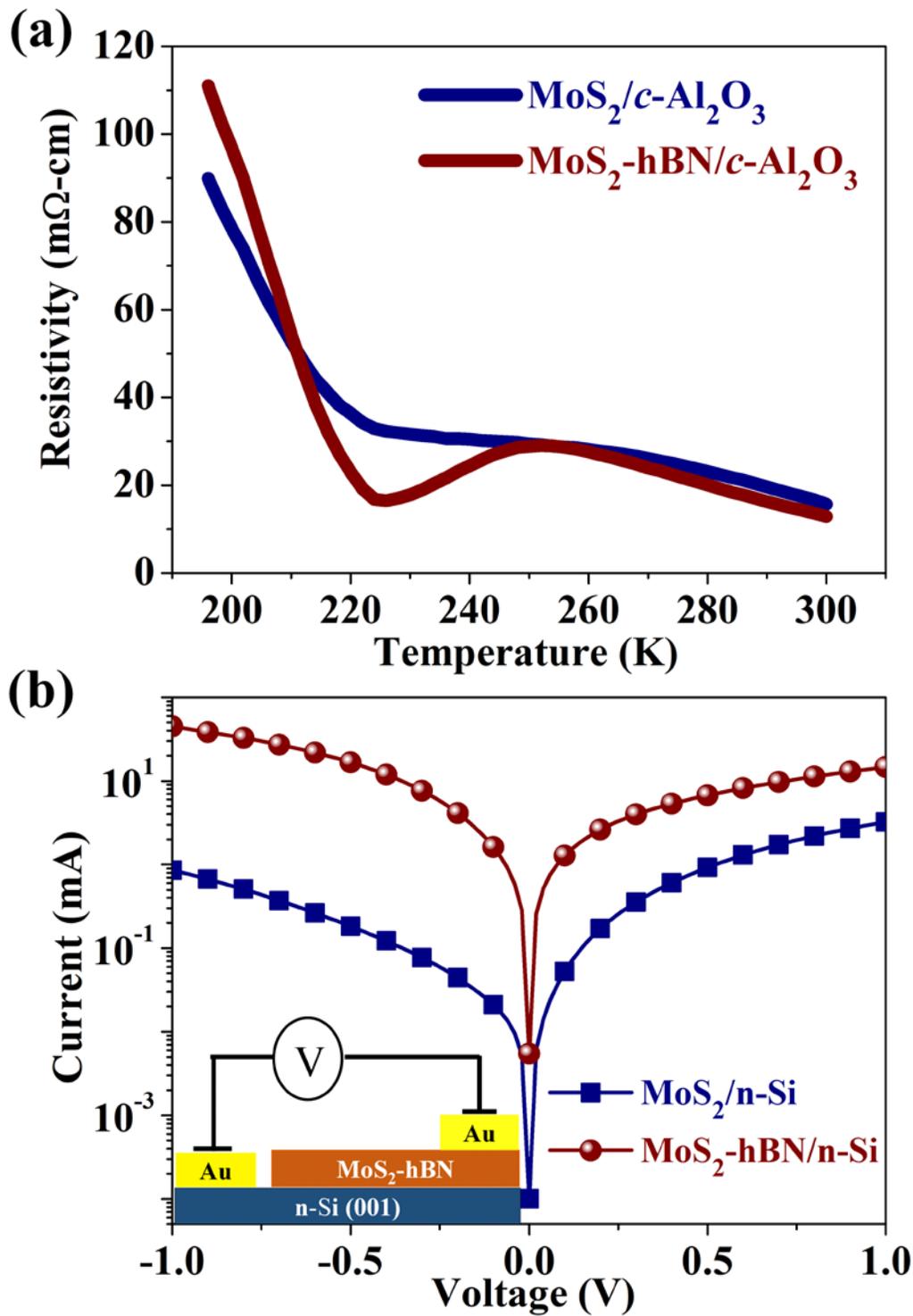

**(Figure-6)**

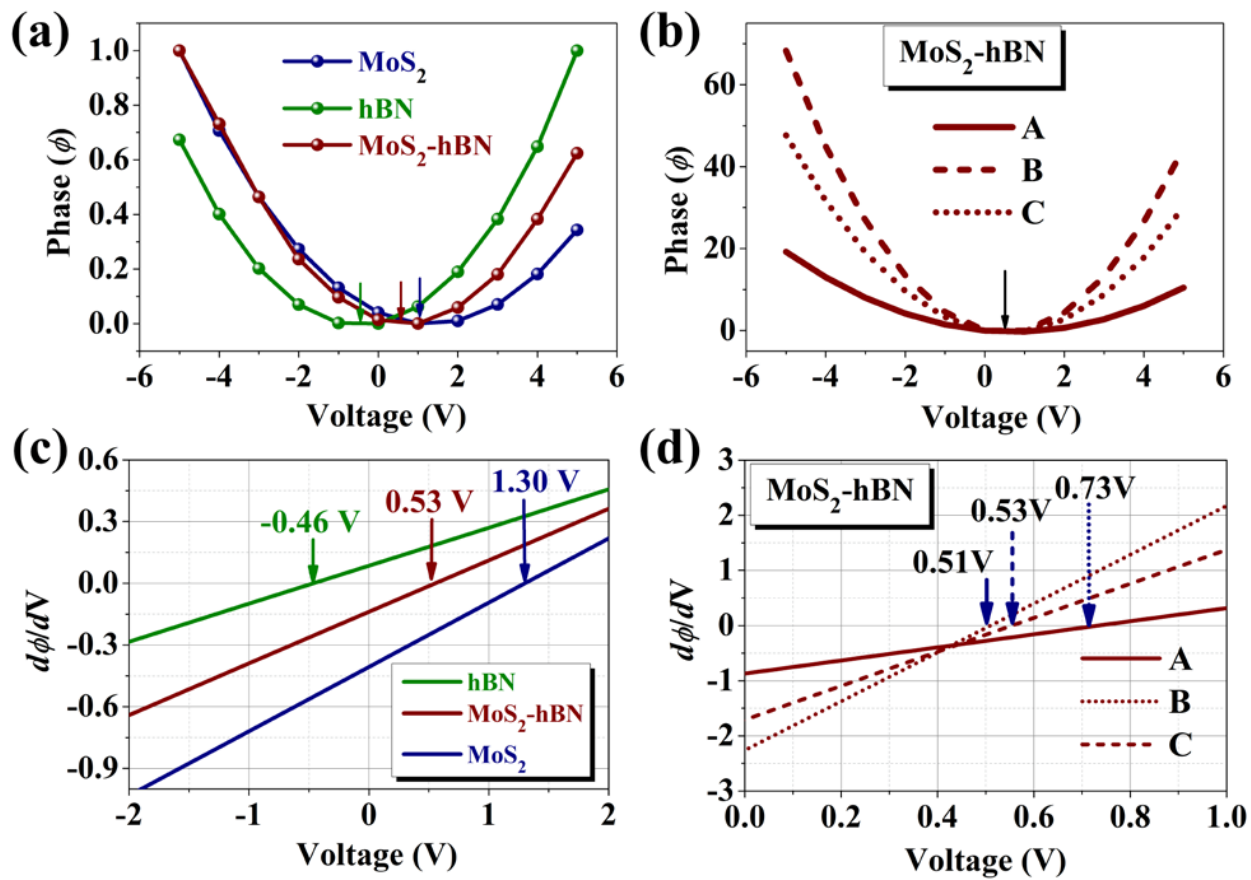

(**Figure-7**)

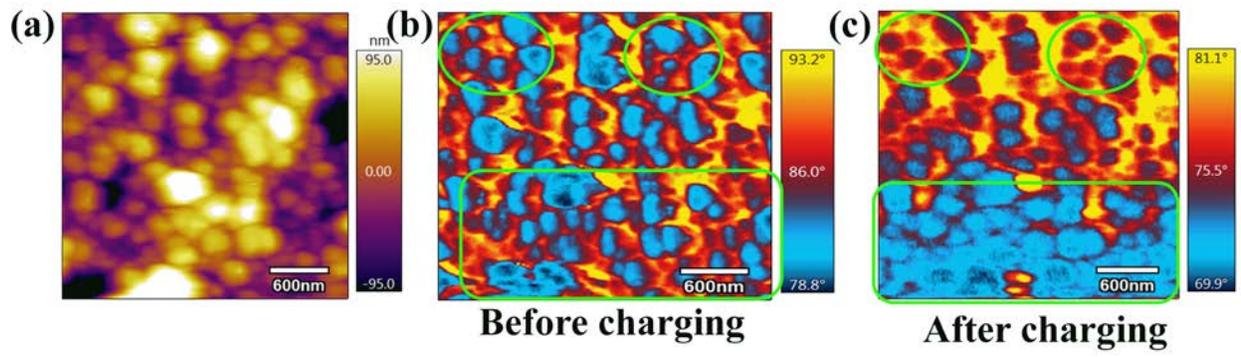

**(Figure-8)**

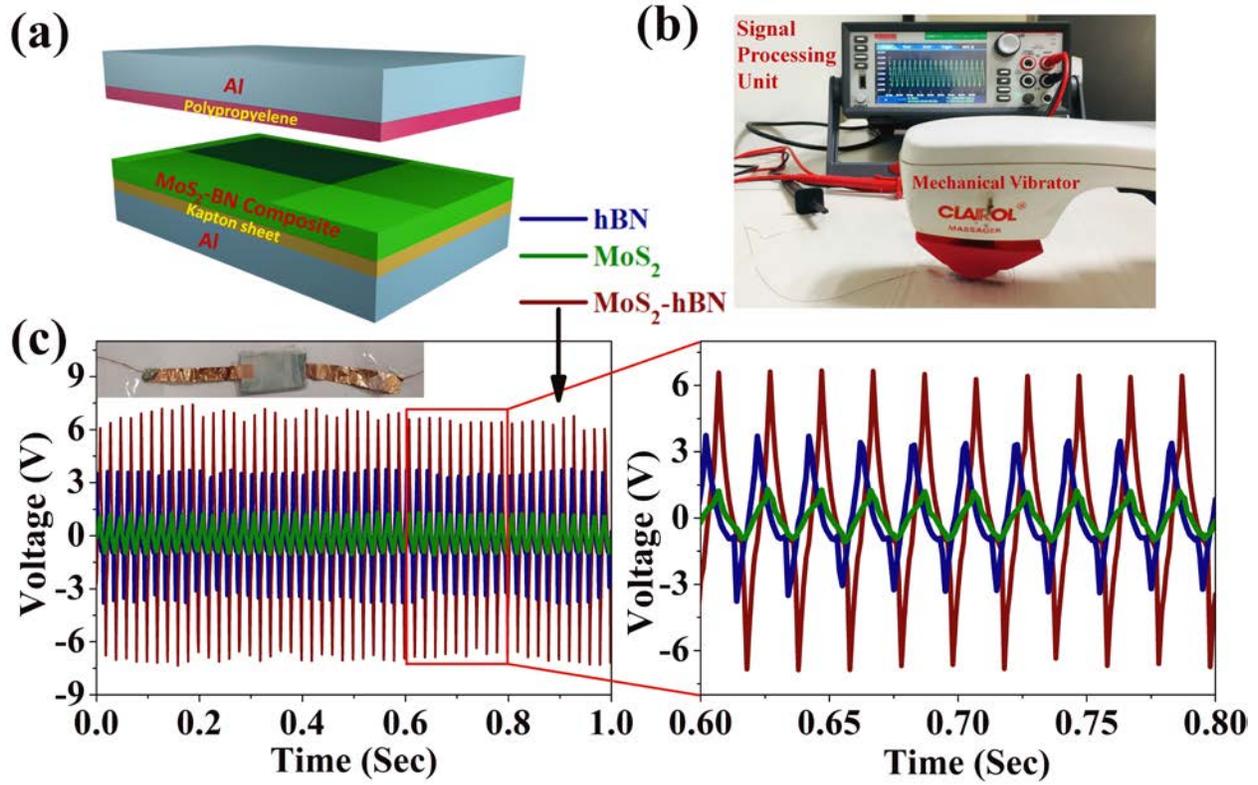

**(Figure-9)**